\documentclass[aps,pra,twocolumn,superscriptaddress,citeautoscript]{revtex4-1}

\usepackage{amsmath,amssymb}
\usepackage{bm}
\usepackage{graphicx}
\usepackage{epstopdf}
\usepackage{latexsym}
\usepackage{subfigure}
\usepackage{color}
\usepackage{dsfont} %for ensemble alphabet, and works with numbers too, use \mathds{1}
\usepackage{wasysym} %for hexagon (in Eq environment or in text)
\usepackage{hyperref} %for links in pdf

\newcommand{\be}{\begin{equation}}
\newcommand{\ee}{\end{equation}}
\newcommand{\bea}{\begin{eqnarray}}
\newcommand{\eea}{\end{eqnarray}}

\begin{document}

\title{A New Type of Quantum Criticality in the Pyrochlore Iridates}

\author{Lucile Savary}
\author{Eun-Gook Moon}
\affiliation{Department of Physics, University of California, Santa Barbara, CA 93106-9530, U.S.A.}
\author{Leon Balents}
\affiliation{Kavli Institute for Theoretical Physics, University of
  California, Santa Barbara, CA 93106-4030, U.S.A.}

\date{February 4, 2014}
\begin{abstract}
Magnetic fluctuations and electrons couple in intriguing ways in the vicinity of zero temperature phase transitions -- quantum critical points -- in conducting materials.  Quantum criticality is implicated in non-Fermi liquid behavior of diverse materials, and in the formation of unconventional superconductors.  Here we uncover an entirely new type of quantum critical point describing the onset of antiferromagnetism in a nodal semimetal engendered by the combination of strong spin-orbit coupling and electron correlations, and which is predicted to occur in the iridium oxide pyrochlores.  We formulate and solve a field theory for this quantum critical point by renormalization group techniques, show that electrons and antiferromagnetic fluctuations are strongly coupled, and that both these excitations are modified in an essential way. This quantum critical point has many novel features, including strong emergent spatial anisotropy, a vital role for Coulomb interactions, and highly unconventional critical exponents.  Our theory motivates and informs experiments on pyrochlore iridates, and constitutes a singular realistic example of a non-trivial quantum critical point with gapless fermions in three dimensions.  
\end{abstract}

\maketitle

Antiferromagnetic quantum critical points (QCPs) are controlled  by the interactions between electrons and magnetic fluctuations \cite{sachdev2011,lohneysen2007fermi}.  In three dimensional metals with a Fermi surface, it is believed to be sufficient to consider Landau damping of the magnetic order parameter in a purely order parameter theory, which leads, following Hertz \cite{hertz1976, millis1993}, to mean field behavior.  In two dimensions, the electronic Fermi surface and order parameter are strongly coupled, a fact which may be related to high-temperature superconductivity and associated phenomena.  This problem is highly non-trivial and still an active research topic \cite{lee2009,metlitski2010quantum,mross2010controlled,efetov2013pseudogap}.

In this paper, we uncover a new antiferromagnetic QCP which is strongly coupled in {\em three} dimensions, engendered by spin-orbit coupled electronic structure.  
We consider a quadratic band-touching at the Fermi energy, as in the inverted band gap material HgTe, but having in mind the strongly correlated family of iridium oxide pyrochlores \cite{wan2011,moon2012,chen2013,kondo2013}. The latter have chemical formula A$_2$Ir$_2$O$_7$, and an antiferromagnetic phase transition indeed occurs both as a function of temperature and at zero temperature with varying chemical pressure (ionic radius of A) \cite{matsuhira2011}.  We show that the replacement of the Fermi surface by a point Fermi node alters the physics in an essential way, suppressing screening of the Coulomb interaction and allowing the order-parameter fluctuations to affect {\em all} the low-energy electrons. These two facts lead to a strongly-coupled quantum critical point.  

\begin{figure}[htbp]
\includegraphics[width=3.3in]{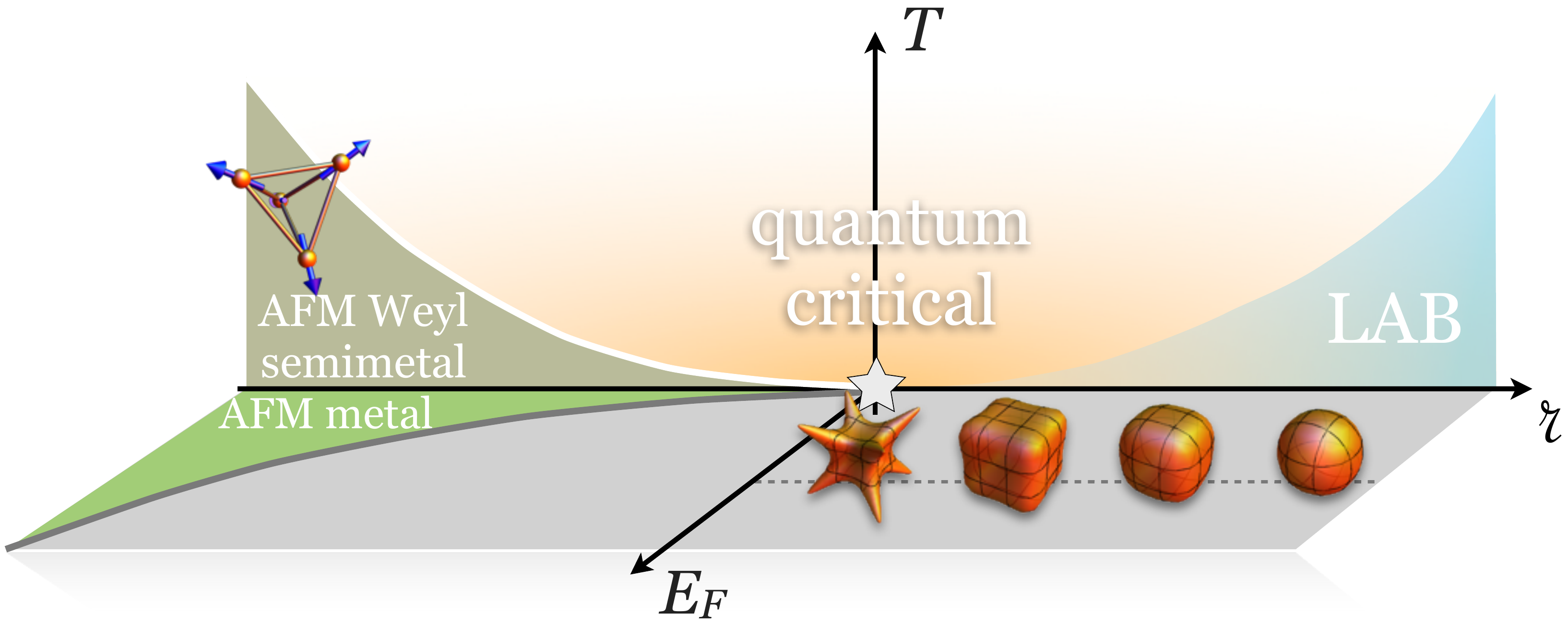}
\caption{Quantum critical point (QCP) and quantum criticality driven by the onset of ``all-in-all-out'' magnetism. For $r\geq r_c$ (in this figure the star indicates $r_c$), the ``Luttinger-Abrikosov-Beneslavskii'' (LAB) phase occurs at $T=0$, with a quadratic Fermi node, while antiferromagnetic (AFM) ``all-in-all-out'' ordering occurs for $r<r_c$, with the quadratic node replaced by linear Weyl points -- a Weyl semimetal. The quantum critical regime occurs at $T>0$ around $r=r_c$.  Note that the quantum critical-AFM boundary (thick white line) is a true (continuous Ising) phase transition. The $E_F$ axis represents the Fermi energy and parametrizes electron or hole doping. The three-dimensional (orange) surfaces represent the shapes of the corresponding Fermi surfaces at small doping -- the increased anisotropy is apparent as one moves towards the QCP. The phase transition denoted by the thick gray line is expected to exhibit critical properties appropriate to a $\mathbf{q}=\mathbf{0}$ order parameter coupled to a Fermi surface, as in the Hertz formulation \cite{hertz1976}, though subject to the usual uncertainties regarding the theory of that problem \cite{lee2009,metlitski2010quantum,mross2010controlled}.  }
\label{fig:QCP}
\end{figure}

The nodal nature of the Fermi point, happily, also enables a rather complete analysis of the problem, which we present here, using the powerful renormalization group (RG) technique.  
The complete theory we present is in sharp contrast to the strongly coupled Fermi surface problem in two dimensions, which remains only partially understood and controversial.  
Finally, the pyrochlore quantum critical point has a remarkable symmetry structure.  We find that, unlike at most classical and quantum phase transitions, rotational invariance is strongly broken in the critical theory: the fixed point ``remembers'' the cubic anisotropy of space (and indeed takes it to an extreme limit, as explained further below).   Compensating for the absence of spatial rotational invariance is, however, an {\em emergent} $SO(3)$ invariance of the critical field theory, which is a purely internal symmetry and unrelated to spatial rotations.   The {\em an}isotropy in real space manifests for example in the formation of ``spiky'' Fermi surfaces when the system close to the QCP is doped with charge carriers, as seen in Fig.~\ref{fig:QCP}.  

To proceed with the analysis, we couple the electrons to an Ising magnetic order parameter $\phi$.  This corresponds for the pyrochlore iridates to the translationally-invariant ``all-in-all-out'' (AIAO) antiferromagnetic state (see ``inset'' in Fig.~\ref{fig:QCP}), for which there is considerable evidence \cite{sagayama2013determination,tomiyasu2012emergence,disseler2012magnetic}.  Due to the time-reversal and inversion symmetries of the paramagnetic state, electron bands are two-fold degenerate, so that band touching necessitates a minimal four-band model.    Therefore the Hamiltonian is expressed in terms of four-component fermion operators $\psi$, $\psi^\dagger$, in addition to $\phi$ and the electrostatic field $\varphi$, which mediates the Coulomb interactions.  The action is 
\begin{eqnarray}
\mathcal{S}&=&\int d^3 x d \tau \psi^{\dagger}(\alpha \partial_{\tau} + \mathcal{H}_0(-i \nabla) +i e \varphi + g M\phi)\psi \nonumber \\
&+& \int d^3 x d \tau \frac{1}{2}\left[{(\nabla\varphi)^2} + {(\nabla \phi)^2} +{(\partial_{\tau} \phi)^2} + r \phi^2 \right],
\label{eq:action}
\end{eqnarray}
where the momentum cutoff ($\Lambda$) is assumed, and where the Hamiltonian density is $\mathcal{H}_0( \mathbf{k})=c_0 \mathbf{k}^2+ \sum_{a=1}^5  \hat{c}_a d_a( \mathbf{k})\Gamma_a$.  Higher-order terms omitted in Eq.~(\ref{eq:action}) prove irrelevant at the QCP. The $d_a$'s (given in the Supp.\ Mat.\ \cite{suppmat}) make a complete basis of the allowed terms quadratic in $k_j$, chosen such that $d_{1,2,3}$ belong to a three-dimensional representation (often called $T_{2g}$) and $d_{4,5}$ make a two-dimensional one (commonly referred to as $E_g$), the $\Gamma_a$'s are anticommuting unit matrices, $\{\Gamma_a,\Gamma_b\}=2\delta_{ab}$, $\Gamma_{ab}=\frac{-i}{2}[\Gamma_a,\Gamma_b]$, $\hat{c}_1=\hat{c}_2=\hat{c}_3=c_1$ and $\hat{c}_4=\hat{c}_5=c_2$ (as they should since they belong to the same representation), and symmetry dictates \cite{suppmat} the order parameter couples via the matrix $M=\Gamma_{45}$ ($\in A_{2g}$). $e$ is the magnitude of the electron charge, and $g\in\mathbf{R}$ parametrizes the coupling strength of the fermions to the order parameter. As discussed in the Supplementary Material \cite{suppmat}, $c_{0,1,2}$ may always be chosen positive, without loss of generality. $c_0$ parametrizes ``particle-hole asymmetry'', with $c_0=0$ denoting a symmetric band structure. Also, when $c_0\leq c_1/\sqrt{6}$, in the vicinity of the Gamma point, the bands touch at and only at the Gamma point. We assume that the system parameters fall within this range, and find that this is consistent.

The model in Eq.~(\ref{eq:action}) has two phases.  For $r>r_c \sim g^2$ (where $r_c$ is thereby defined), $\phi$ fluctuates around zero, and can be integrated out.  This is a magnetically disordered state.  The resulting model with Coulomb interactions alone describes a non-Fermi liquid {\em phase}, as first discussed by Abrikosov and Beneslavskii \cite{abrikosov1971,abrikosov1974}  and thoroughly revisited recently \cite{moon2012}.   Notably, in this regime, non-trivial scaling exponents arise and the low-energy electronic dispersion renormalizes to become {\em isotropic}, i.e. effectively $c_1 \rightarrow c_2$ and $c_0 \ll c_1$.   For $r<r_c$, the expectation value $\langle\phi\rangle \neq 0$, and replacing $\phi \rightarrow \langle \phi\rangle$ causes the two-fold degenerate bands to split, removing the quadratic touching at $\mathbf{k}=\mathbf{0}$ in favor of eight linearly-dispersing ``Weyl points'' along the $\langle 111\rangle$ directions: a {\em Weyl semimetal}.

\begin{figure}[htbp]
\includegraphics[width=1.8in]{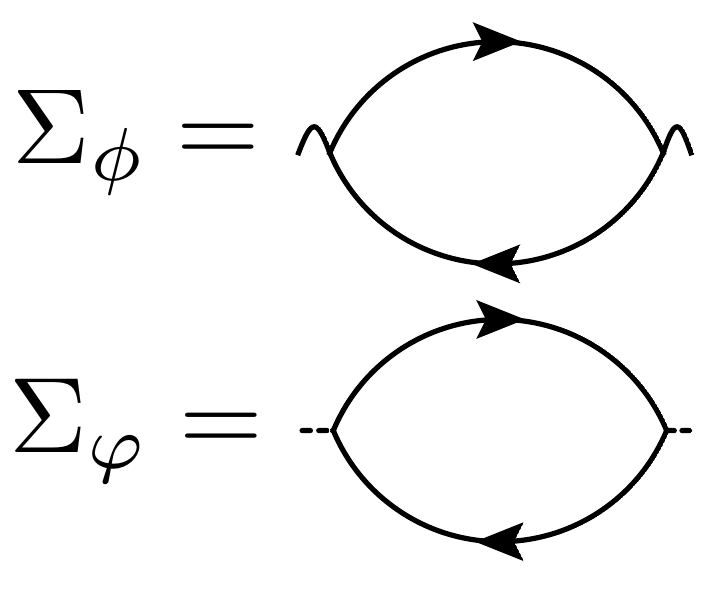}
\caption{Boson self energies for the order parameter ($\Sigma_\phi$) and electrostatic ($\Sigma_\varphi$) fields.}
\label{fig:bubble}
\end{figure}

\begin{figure}[htbp]
\includegraphics[width=3.3in]{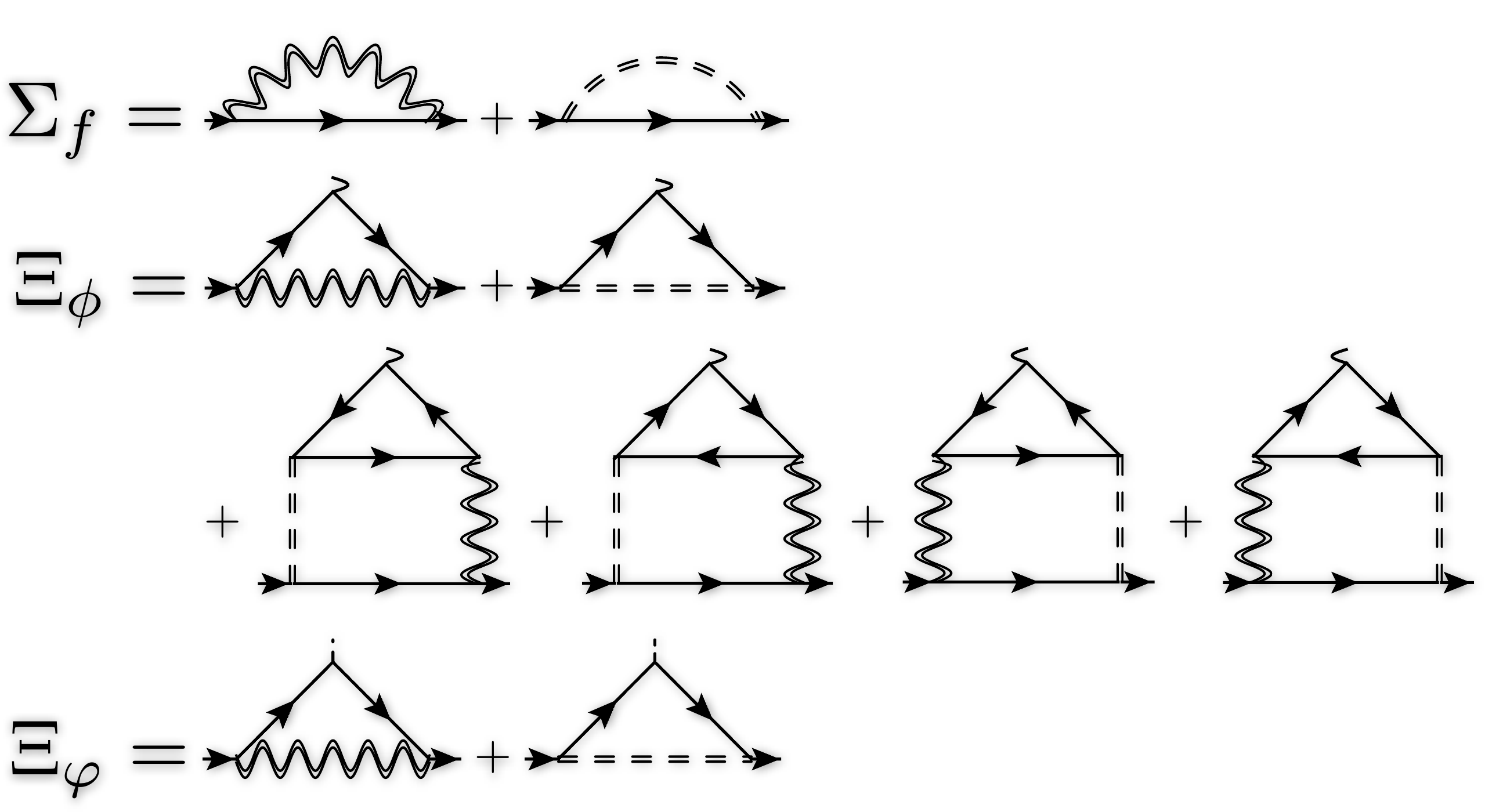}
\caption{$1/N$ diagrams for the fermion self-energy $\Sigma_f$ and vertex corrections $\Xi_\phi$ and $\Xi_\varphi$ (only two-loop diagrams that need be calculated, i.e.\ that do not vanish or cancel one another, are shown).  Double lines indicate the renormalized boson propagators including the self-energies from Fig.~\ref{fig:bubble}. Expressions for these diagrams are given in the Supplementary Material \cite{suppmat}.}
\label{fig:1oNdiags}
\end{figure}

We now turn to the critical regime. To proceed, we introduce as a formal device $N$ copies of the four fermion fields, replacing $g\rightarrow g/\sqrt{N}$ (resp.\ $ie\rightarrow ie/\sqrt{N}$) and $\Gamma_a \rightarrow \Gamma_a \otimes {\sf 1}_N$ (${\sf 1}_N$ is the $N\times N$ identity matrix).  We organize perturbation theory in powers of $1/N$, but in the end argue that the results are asymptotically exact for the physical case $N=1$.   To leading order in $1/N$, we require the two boson self-energies in Fig.~\ref{fig:bubble}, and, using the dressed boson propagators including this correction, the fermion self-energy and vertex functions in Fig.~\ref{fig:1oNdiags}.  These diagrams allow a full calculation of the $O(1/N)$ terms of all critical exponents.   The evaluation of the diagrams is complicated by the three mass parameters of the free fermion propagators.  Fortunately, a simplification is possible due to the structure of the RG.  While the (inverse) mass terms $c_0$, $c_1$, and $c_2$ all have identical engineering dimensions, they, in general renormalize differently from loop corrections, and thus their ratios {\em flow} in the full RG treatment.  We find below that, in the critical regime, $c_0/c_2, c_1/c_2\rightarrow 0$ under renormalization (arguments why this is the only reasonable choice are given in the Supplementary Material \cite{suppmat}).  This allows technical simplifications in the loop integrals, and also has physical consequences we explore later.

In particular, in the limit $c_1/c_2 \rightarrow 0$, the interband splitting vanishes along the $\langle 111\rangle$ directions, leading to an extended singularity of the electron Green's function.   In the loop integrals determining the bosonic self-energies, this produces a divergent contribution at non-zero ${\bf k}$.  Technically, with the assumptions $c_0/c_2, c_1/c_2 \ll 1$ and $c_0/c_1 < 1/\sqrt{6}$ (shown self-consistent below), the low-energy behavior (small $\omega_n,\mathbf{k}$) may be extracted as (see Supplementary Material \cite{suppmat})
\begin{equation}
\Sigma_b(\omega_n,\mathbf{k})= - r_{b}^{c}+\frac{g_b^2}{\alpha}\left(|\ln\,c_1/c_2\,||\mathbf{k}|f_b(\mathbf{\hat{k}})+\sqrt{|\omega_n|}C_b\right),
\label{eq:sigmab}
\end{equation}
where $r_\phi^c = r_c \sim g^2\Lambda$, where $\Lambda$ is an upper momentum cutoff, $g_\phi=g$, $g_\varphi=ie$, and  $r_\varphi^c=C_\varphi=0$ follows from charge conservation.  $C_\phi\approx 1.33$ %hellonumbers 
and the functions $f_b(\mathbf{\hat{k}})$ are given as integrals in the Supplementary Material \cite{suppmat}.   

Note that, at low energy, the dispersive terms in Eq.~(\ref{eq:sigmab}) are much larger than the bare $\mathbf{k}^2,\omega_n^2$ terms they correct, and hence dominate the renormalized Green's functions. Thus, in the fermion self-energy and vertex correction, the renormalized boson propagator, $\mathcal{G}_b^{-1}=\mathcal{G}_{b;0}^{-1}+\Sigma_b\approx\Sigma_b+r_b$ (note $r_\varphi=0$), must be used. 
This renormalized boson propagator corresponds to the $N=\infty$ result, and already reveals some dramatic features.  First, the bosons immediately receive a large anomalous scaling dimension, equal to $1$, and their dynamics becomes damping-dominated, with dynamical critical exponents close to $2$.  Second, since the damping terms which dominate $\mathcal{G}^{-1}_b$ are proportional to $g_b^2$, it implies that the fermion self-energies, which involve two interaction vertices (see Fig.~\ref{fig:1oNdiags}), become $g_b$ independent: this is a sign of universality at the QCP.  

To confirm the assumed scaling of $c_1/c_2$, $c_0/c_2$, and fully determine the critical behavior, we turn to the renormalization group approach. There, as usual, we apply the following rescaling (applicable in real space)
\begin{eqnarray}
&&x\rightarrow e^\ell x,\quad\tau\rightarrow e^{\int_0^\ell d\ell' z(\ell')}\tau,\quad\psi\rightarrow e^{-\int_0^\ell d\ell'\Delta_\psi(\ell')}\psi,\nonumber\\
&&\phi\rightarrow e^{-\int_0^\ell d\ell'\Delta_\phi(\ell')}\phi,\quad\varphi\rightarrow e^{-\int_0^\ell d\ell'\Delta_\varphi(\ell')}\varphi,
\end{eqnarray}
where $\ell\geq0$ parametrizes the RG flow. The exponents are left allowed to be scale dependent, as is necessary \cite{huh2008}, as we shall see below. 

We evaluate the contributions to the fermion propagator and coupling constants due to a small change in the cutoff (which corresponds physically to integrating out modes to keep the rescaled cutoff unchanged).  Hence, the RG flow equations are obtained by {\em (i)} logarithmically differentiating the fermion self energy and vertex functions with respect to the cutoff $\Lambda$ (made soft through a rapidly decaying function $|\mathbf{q}|/\Lambda\mapsto\mathcal{F}(|\mathbf{q}|/\Lambda)$) \cite{huh2008,vojta2000a}, and {\em (ii)} identifying the appropriate coefficients of the Taylor expansion (in $k_i$ and $\omega_n$) of the result \footnote{Note that a number of technicalities are involved in this calculation, in particular regarding the convergence of the differentiated functions; All are discussed in the Supplementary Material \cite{suppmat}.}.

We leave most details to the Supplementary Material \cite{suppmat}, and only give one example here.  To extract the correction to the mass coefficient $c_1$, we first expand the fermion self-energy as
\begin{equation}
\Sigma_{f}(\omega_n,\mathbf{k})=\Sigma_{f}^0 I+\sum_{a=1}^5\Sigma_{f}^a\Gamma_a,
\end{equation}
and examine the $\Sigma_f^1$ component.  The RG equation is then
\begin{equation}
\frac{\partial_\ell c_1}{c_1}=z+1-2\Delta_\psi+\frac{\sqrt{2}}{c_1}\left.\left(\partial_{k_x,k_y}^2\left[\Lambda\frac{d}{d\Lambda}\Sigma_f^1\right]\right)\right|_{\omega_n=0,\mathbf{k}=\mathbf{0}}
\label{eq:floweg}
\end{equation}
(we define $d_1(\mathbf{k})=k_xk_y/\sqrt{2}$ \cite{suppmat}). Similar expressions are obtained for the other parameters of the theory, $c_0$, $c_2$, $\alpha$, $g$ and $ie$.  The latter all depend on $c_1$ and $c_2$ through $1/(N|\ln c_1/c_2\,|)$ or $1/(N|\ln c_1/c_2\,|^2)$ (expressions are expanded in small $1/|\ln c_1/c_2\,|$, see Supp.\ Mat.\ \cite{suppmat}).  Therefore, for the six equations thereby obtained, there are four unknowns ($z$, $\Delta_\psi$, $\Delta_\phi$ and $\Delta_\varphi$) which can be chosen to keep four parameter fixed, leaving two left to flow. Here we find it is possible to keep $\alpha$, $g$, $(ie)$ and $c_2$ fixed, and thus $c_1/c_2$ and $c_0/c_2$ will flow.  Note that, in doing so, we obtain a critical theory with non-zero coupling of fermions both to order-parameter and Coulomb-potential fluctuations: both effects are crucial and important in stabilizing the QCP.  Finally, we obtain
\begin{equation}
z(\ell)=2-\delta z(\ell),\; \Delta_\psi(\ell)=\frac{3+\eta_\psi(\ell)}{2},\; \Delta_b(\ell)=\frac{3+\eta_b(\ell)}{2},\label{eq:1}
\end{equation}
%hellonumbers
where $\delta z=\frac{0.0634}{|\ln c_1/c_2\,|N}$, $\eta_\psi=\frac{0.287}{|\ln c_1/c_2\,|^2N}$, $\eta_\phi=1+\frac{0.510}{|\ln c_1/c_2\,|N}$ and $\eta_\varphi=1-\frac{0.127}{|\ln c_1/c_2\,|N}$. 

The flow equations may be solved thanks to that of $c_1/c_2$, which is an analytically-soluble differential equation involving only $c_1/c_2$ \cite{suppmat}. Ultimately, we find
\begin{equation}
(c_1/c_2)(\ell)=e^{-\frac{\upsilon_0}{\sqrt{N}}\sqrt{\ell+\ell_0}},\; \mbox{and}\; (c_0/c_1)(\ell)=\Upsilon_0 e^{-\frac{\upsilon_0'}{\sqrt{N}}\sqrt{\ell+\ell_0}},\label{eq:2}
\end{equation}
%hellonumbers
with $\upsilon_0=0.202$, $\upsilon_0'=0.424$ and where $\ell_0$ and $\Upsilon_0$ are constants which depend on the system's parameters, namely on $c_{0,1,2}(\ell=0)$.   Formally, therefore both the $c_0$ and $c_1$ mass terms are irrelevant in the RG sense, but they can be ``dangerously irrelevant'' insofar as they control certain physical properties (see below).  Note also that not only is $c_0$ irrelevant, but it also flows to zero {\em faster} than $c_1$, so that $c_0/c_1$ becomes small at the QCP.  

Intuition for the irrelevance of $c_1$ comes from considering the fermion self-energy $\Sigma_f$, which yields the corrections to $c_{0,1,2}$ and to $\alpha$, and is given {\em schematically} by $\Sigma_f=\mathcal{G}_\phi MG_0M+\mathcal{G}_\varphi G_0$ (the contributions from each boson field just add up). In the first term, which represents dressing of electrons by magnetic fluctuations, the appearance of $M$, which commutes with $\Gamma_{1,2,3}$ but {\em anti-}commutes with $\Gamma_{4,5}$, portends ``opposite'' consequences for $c_1$ and $c_2$.   The second term, due to Coulomb effects, tends instead to affect $c_1$ and $c_2$ identically.  Our calculation shows that the former tendency prevails, and $c_1/c_2 \rightarrow 0$ under RG, as claimed above.  Conversely, the fact that $c_0/c_1 \rightarrow_{\ell\rightarrow \infty} 0$ should be attributed to the effect of Coulomb forces, which suppress particle-hole asymmetry.  Indeed, we have checked that if in the calculations we artificially turn off the long-range Coulomb potential, i.e. take $e=0$, the QCP is unstable and there is no direct, continuous quantum phase transition from the LAB state to the AIAO one\,\cite{suppmat}.

Eqs.~(\ref{eq:1},\ref{eq:2}) determine the properties at the QCP.  We now turn to a discussion of the physical consequences.  First we consider some scaling properties.  For the  correlation length, we need the flow equation for $\delta r=r-r_c$, the deviation from the critical point: $\partial_\ell (\delta r)=\nu^{-1} (\delta r)$, with $\nu = 1/[2-\eta_\phi(\ell)-\delta z(\ell)]$.  This implies, in the usual way, that the correlation length behaves as $\xi \sim (\delta r)^{-\nu}$, up to logarithmic corrections.  Also interesting is the order parameter growth in the AIAO phase.  By scaling, $\langle \phi\rangle \sim \xi^{-\Delta_\phi} \sim |\delta r|^\beta$, with $\beta = \Delta_\phi \nu$.   We also expect the critical temperature of the magnetic state to obey $T_{c} \sim \xi^{-z} \sim |\delta r|^{z\nu}$.   In asymptopia, i.e. $\ell \rightarrow \infty$, all the $N$-dependent corrections vanish, and the exponents correspond to those of a saddle-point treatment of $\varphi,\phi$.  These are still distinct from the usual order parameter mean field theory, as witnessed by the large ($\eta_\phi^\infty=\eta_\varphi^\infty=1$) anomalous dimensions in this limit, and the unconventional values $\nu^\infty=1$, $\beta^\infty= (z\nu)^\infty = 2$.   The latter is noteworthy insofar as it implies an unusually wide critical fan at $T>0$ which is controlled by the QCP (see Fig.~\ref{fig:QCP}).  The RG treatment goes beyond the saddle point in giving the corrections due to finite $c_1/c_2$, which are small only logarithmically, and thus may be significant for physically-realistic situations. For example we find $\langle \phi\rangle\sim (\delta r)^2\exp\left[\frac{13.9}{\sqrt{N}}\sqrt{\ln\frac{\delta r}{r_0}}\right]$ \cite{suppmat}, where $r_0$ is a constant.
%hellonumbers 2(5aphi+2az)/upsilon0

The irrelevance of $c_0$ and $c_1$ has other, more direct, physical consequences.  Because of the former, the low-energy electronic spectrum becomes approximately particle-hole symmetric.  The latter has more implications.  Obviously, the electronic spectrum develops pronounced cubic anisotropy, with anomalously low energy excitations along the cubic $\langle 111\rangle$ directions in momentum space.  This is in stark contrast to most critical points (for example of Ginzburg-Landau type, or involving Dirac fermions), which typically have emergent spatial isotropy and even conformal symmetry and Lorentz invariance at the fixed point. 
These low-energy excitations manifest, for example, in the specific heat $c_v$.  Since at the Gaussian level the coefficient of $T^{3/2}$ {\em diverges} as $c_1^{-3/2}$, we estimate, by using $\ell \sim \frac{1}{z}\ln T_0/T$ as a cut-off ($T_0$ is a microscopic energy scale),  $c_v\sim \exp\left[\frac{3\upsilon_0}{2\sqrt{N}\sqrt{z}}\sqrt{\ln \frac{T_0}{T}}\right]T^{3/2}$, with $z\approx2$ \cite{suppmat}.  The emergent anisotropy may also manifest in increasingly-``spiky'' Fermi surfaces in lightly doped samples near the QCP, see Fig.~\ref{fig:QCP}.  

 Although rotational symmetry is strongly broken, the vanishing of $c_1$ leads to an {\em emergent internal} $SO(3)$ symmetry, corresponding to rotating the $\Gamma_a$ matrices with $a=1,2,3$ amongst themselves like a vector.   The generator of this symmetry is the $SU(2)$ pseudo-spin ${\bf I}$, with
\begin{equation}
  \label{eq:3}
  I_a = -\frac{1}{4}\epsilon_{abc} \, \psi^{\dagger} \Gamma_{bc}\psi  = \psi^{\dagger} \left(-\frac{7}{6} J_a + \frac{2}{3} J_a^3\right)\psi,
\end{equation}
where $a=x,y,z=1,2,3$.  Its integral has $SU(2)$ commutation relations and commutes with the fixed-point Hamiltonian.

{\em Discussion.---} In standard Hertz-Millis theory \cite{hertz1976, millis1993}, the inequality $d+z>4$ implies that the theory is above its critical dimension, and thus has mean field behavior. Although this inequality holds here, taking $z=2$, the conclusion is false.  The Hertz-Millis approach assumes the fermions may be innocuously integrated out, and obtains this inequality by power-counting the $\phi^4$ term in the Landau action, which is irrelevant.  Instead, here we have strong coupling of fermions with the order parameter, and the coupling term $\sim \phi \psi^\dagger \psi$ is {\em marginal} using $z=2$, $\Delta_\phi = 2$, $\Delta_\psi=3/2$.  If one {\em does} integrate out the fermions, one obtains a nonanalytic $|\phi|^{5/2}$ term \cite{suppmat}, which overwhelms the na\"ive $\phi^4$ one, and is again marginal by power counting.   This $|\phi|^{5/2}$ dependence was obtained previously in Ref.~\cite{kurita2013}, in the context of a mean-field treatment of related transitions.  Note, however, that such a mean-field analysis integrating out fermions is not justified and misses important physics.

Our critical theory has some formal similarity to the theory of a two-dimensional nodal nematic QCP in a $d$-wave superconductor \cite{huh2008}, insofar as both theories display ``infinite anisotropy'': in our case due to $c_1/c_2 \rightarrow 0$ under RG.  This suggests that, as in Ref.~\cite{huh2008}, at low energy {\em the perturbative expansion parameter is small for all $N$}, and that therefore our results apply directly at low energy to the physical case $N=1$.  This conclusion is appealing, though we have not shown it rigorously.

With the above results in hand, we comment on the connection to experiments.  In the pyrochlore iridates, the QCP might be tuned by alloying the A-site atoms, e.g.  Pr$_{2-2x}$Y$_{2x}$Ir$_{2}$O$_7$, or by pressurizing stoichiometric compounds nearby.  The theory developed here, which relies only on cubic symmetry and strong SOC, may apply to other materials if the bands at the Fermi energy belong to the appropriate irreducible representation, and it would be interesting to search for other examples. Experimentally, the heavily-damped paramagnon could be observed in inelastic neutron or x-ray scattering.  An explicit calculation of the fermion spectral function measured in angle resolved photoemission has been made neither here nor for the non-Fermi liquid paramagnetic state \cite{moon2012}, and is an important problem for future theory.  However, in general, the weak logarithmic flow of the Hamiltonian parameters signifies large self-energy corrections, and behavior somewhat similar to marginal Fermi-liquid theory may be expected.  

We also mention some possible complications in the iridates.  Impurity scattering is a relevant perturbation and hence important at low energy close to the band touching. Therefore, our results will apply best in the cleanest samples.  Also, an accidental band crossing may occur away from the zone center, thereby shifting the Fermi level a few meV away from the nodal point.  This should be addressed by {\em ab initio} calculations and experiments.  In such a case, our results still hold for energies and/or temperatures above this shift energy. Finally, in many of the pyrochlore iridates, the A-site ion hosts rare-earth moments, which were not included here.  They only weakly couple to the Ir electrons and to themselves, so are only important at low energy.   On the antiferromagnetic side of the QCP, the Ir spins act as strong local effective magnetic fields, locking the A-site spins. However, when the Ir sites are not ordered, as in Pr$_2$Ir$_2$O$_7$, A-site ions will have an effect below a few Kelvins. Several authors have proposed scenarios based on RKKY interactions \cite{chen2012,flint2013,lee2013}, but the quantum critical theory expounded here should be an apt starting point for a systematic analysis.

{\em Acknowledgements.---} We thank Cenke Xu and Yong-Baek Kim for discussions on prior
  work, Max Metlitski for pointing Ref.~\cite{huh2008} to us,
  and acknowledge Ru Chen, Satoru Nakatsuji and Takeshi Kondo for
  sharing unpublished data. The integrals were performed using the
  Cuba library \cite{hahn2005}, and the Feynman diagrams in
  Figs.~\ref{fig:bubble} and \ref{fig:1oNdiags} drawn with JaxoDraw
  \cite{binosi2004}.  L.S.\ and L.B.\ were supported by the DOE
  through grant DE-FG02-08ER46524, and E.-G.\ M.\ was supported by the
  MRSEC Program of the National Science Foundation under Award No. DMR
  1121053.  

\bibliography{arxiv.bib}

\vspace{-0.3cm}

\begin{widetext}

\vspace{-0.4cm}

\section*{\large SUPPLEMENTARY MATERIAL}

In reciprocal space, the action, Eq.~(1) in the main text, is
\begin{eqnarray}
\mathcal{S}&=&\int_{-\infty}^{\infty}\frac{d\omega_n}{2\pi}\int_\Lambda\frac{d^3k}{(2\pi)^3}\left[\phi_{-\omega_n,-\mathbf{k}}\left(\frac{\omega_n^2}{2}+\frac{\mathbf{k}^2}{2}+\frac{r}{2}\right)\phi_{\omega_n,\mathbf{k}}+\varphi_{-\omega_n,-\mathbf{k}}\left(\frac{\mathbf{k}^2}{2}\right)\varphi_{\omega_n,\mathbf{k}}\right.\nonumber\\
&&\qquad+\; \psi^{\dagger}_{\omega_n,\mathbf{k}}\left(-\alpha\, i\omega_n+c_0\mathbf{k}^2+\sum_{a=1}^5\hat{c}_ad_a(\mathbf{k})\Gamma_a\right)\psi_{\omega_n,\mathbf{k}}\\
&&\qquad\left.+\;g\int_{-\infty}^{\infty}\frac{d\omega_n'}{2\pi}\int_\Lambda\frac{d^3k'}{(2\pi)^3}\phi_{\omega_n'-\omega_n,\mathbf{k}'-\mathbf{k}} \psi^{\dagger}_{\omega_n',\mathbf{k}'}M\psi_{\omega_n,\mathbf{k}}+ie\int_{-\infty}^{\infty}\frac{d\omega_n'}{2\pi}\int_\Lambda\frac{d^3k'}{(2\pi)^3}\varphi_{\omega_n'-\omega_n,\mathbf{k}'-\mathbf{k}} \psi^{\dagger}_{\omega_n',\mathbf{k}'}\psi_{\omega_n,\mathbf{k}}\right],\nonumber
\label{eq:actionrecipr}
\end{eqnarray}
\end{widetext}
where all the notations are defined in Sec.~\ref{sec:notations} of the
present Supplementary Material. Throughout the latter, for ease of presentation, we shift the QCP so that $r_c=0$.

\section{Notations and symmetries}
\label{sec:notations}

In this section, we provide more information about the notations used
in the main text and a more detailed discussion of the symmetries at play.

\subsection{Fermion Hamiltonian}
\label{sec:fermion}

The fermionic Hamiltonian density in the disordered
(quadratic band touching) phase reads
\begin{eqnarray}
\mathcal{H}_0(\mathbf{k})&=&\alpha_1 \mathbf{k}^2 + \alpha_2 \left(\mathbf{k} \cdot \mathbf{J}\right)^2 + \alpha_3 \left(k_x^2 J_x^2 + k_y^2 J_y^2 +k_z^2 J_z^2\right) \nonumber \\
&=&c_0 \mathbf{k}^2+ \sum_{a=1}^5  \hat{c}_a d_a( \mathbf{k})\Gamma_a, 
\end{eqnarray}
where $\hat{c}_1=\hat{c}_2=\hat{c}_3=c_1$ and $\hat{c}_4=\hat{c}_5=c_2$.
The first line uses the conventional Luttinger parameters
($\alpha_{1,2,3}$) in the $j=3/2$ matrix representation \cite{luttinger1956}, and the
second line is the form used in the main text. The Gamma
matrices ($\Gamma_{a}$) form a Clifford algebra,
$\{\Gamma_a,\Gamma_b\}=2\delta_{ab}$, and have been introduced as described in the literature
\cite{murakami2004}. Note that $c_0$ quantifies the particle-hole asymmetry, while $\left|c_1-c_2\right|$ naturally characterizes the cubic anisotropy. 
The energy eigenvalues are $\mathsf{E}_{\pm}(\mathbf{k}) = c_0 \mathbf{k}^2 \pm
E(\mathbf{k})$, where $E(\mathbf{k})=\sqrt{\sum_{a=1}^5 \hat{c}_a^2  d_a
  ^2( \mathbf{k}) }$ and
\begin{eqnarray}
&&d_1( \mathbf{k})=\frac{k_xk_y}{\sqrt{2}},\quad d_2( \mathbf{k})=\frac{k_xk_z}{\sqrt{2}},\quad d_3( \mathbf{k})=\frac{k_yk_z}{\sqrt{2}} \nonumber \\
&&d_4( \mathbf{k})=\frac{k_x^2-k_y^2}{2\sqrt{2}},\quad d_5( \mathbf{k})=\frac{2k_z^2 -k_x^2-k_y^2}{2\sqrt{6}}. \nonumber 
\end{eqnarray}
It is very important to note that, in the limit $c_{0,1} \rightarrow 0$,
$E(\mathbf{k})$ and 
the energy spectrum $\mathsf{E}_\pm(\mathbf{k})$ become
gapless along the $\langle111\rangle$ directions. When needed, a ``regularization'' is
then possible, for example by introducing higher momentum dependence in $c_{1,2}$, e.g. $c_{1,2}\rightarrow c_{1,2}+\lambda \mathbf{k}^2$. 

It is straightforward to relate the coefficients used in the main text to
the Luttinger $\alpha_i$ parameters. This can be done by expressing
the spin operators in terms of the Gamma matrices, using for example the equalities
\begin{eqnarray}
&& J_x = \frac{\sqrt{3}}{2} \Gamma_{15} - \frac{1}{2} (\Gamma_{23} - \Gamma_{14}) \ ,\nonumber \\
&& J_y = -\frac{\sqrt{3}}{2} \Gamma_{25} + \frac{1}{2} (\Gamma_{13} + \Gamma_{24}) \ ,\nonumber \\
&& J_z = - \Gamma_{34} - \frac{1}{2} \Gamma_{12} \ ,
\end{eqnarray}
where $\Gamma_{ab} = \frac{1}{2 i}
[\Gamma_a, \Gamma_b]$.

The fermion bare Green's function is 
\begin{eqnarray}
G^0_{\omega_n,\mathbf{k}} = \frac{1}{-i\alpha\, \omega_n +\mathcal{H}_0(\mathbf{k})} = \frac{1 }{-i\alpha\, \omega_n + \mathsf{E}_{\epsilon}(\mathbf{k})}  {\rm P}_{\epsilon}(\mathbf{k}),\nonumber 
\end{eqnarray}
where the sum over $\epsilon=\pm1$ is implicit and ${\rm P}_{\epsilon}(\mathbf{k}) = \frac{1}{2}\left(1+\epsilon
\frac{\mathcal{H}_0(\mathbf{k})-c_0\mathbf{k}^2}{E(\mathbf{k})}\right)$
is a projection operator, ${\rm P}_{\epsilon}^2(\mathbf{k})=1$.

\subsection{Symmetries}

It is useful to recap the symmetries of the system in the absence of
all-in-all-out order, and detail the remaining symmetries in its presence.

As defined above and in Refs.~\cite{murakami2004} and
\cite{moon2012}, the $\Gamma_a$ matrices are even under
time-reversal and inversion symmetry, while the $\Gamma_{ab}$ are
even under inversion, but 
odd under time-reversal. 

As is well-known for some semiconductors, like HgTe, the touching of four bands at the Gamma point is protected by
cubic symmetries (the bands at the Gamma point belong to a four-dimensional
representation of the cubic group $O_h$), and the absence of a linear term
follows from time-reversal and cubic (inversion) symmetries. Moreover, thanks to inversion
and time-reversal symmetries, all bands are
doubly-degenerate away from the Gamma point.

The magnetic order parameter field $\phi$ transforms as follows under the
symmetries of the ``disordered'' system. It is odd under time-reversal
symmetry (since the spins
$\vec{\mathsf{S}}\rightarrow-\vec{\mathsf{S}}$ under time-reversal), and so only the (time-reversal-odd) $\Gamma_{ab}$ can couple
to it. It is even under inversion (since 
$\vec{\mathsf{S}}\rightarrow\vec{\mathsf{S}}$ under
inversion), unchanged under three-fold rotations, and
odd under the allowed reflections of the pyrochlore lattice. A single
Gamma matrix, namely $\Gamma_{45}\propto J_xJ_yJ_z+J_zJ_yJ_x$ \cite{murakami2004,moon2012} (see below), transforms
identically. 

The Hamiltonian at {\em fixed} $\mathbf{k}$, i.e.\
$\mathcal{H}_0(\mathbf{k})$, together with the coupling to the order
parameter with $\phi\neq0$, which we call $\mathcal{H}_1(\mathbf{k})$, have the following transformation
properties. For $\mathbf{k}\parallel\langle111\rangle$,
$\mathcal{H}_1$ is invariant under three-fold rotations about
$\mathbf{k}$, and reflections with respect to planes that contain
$\mathbf{k}$. For $\mathbf{k}=\mathbf{0}$, there is additionally
inversion symmetry. The symmetry group at $\mathbf{k}=\mathbf{0}$ then
decomposes the four bands of interest into two two-dimensional
representations. For $\mathbf{k}\neq\mathbf{0}$, symmetries do not
impose bands to cross, hence making any crossings ``accidental.''
However, it is noteworthy that the purely quadratic Hamiltonian $\mathcal{H}_0$ we study,
with $c_0\leq c_1/\sqrt{6}$ and $c_1\leq c_2/\sqrt{6}$, in the
presence of the linear coupling to the order parameter
$\phi\psi^\dagger\Gamma_{45}\psi$ leads inevitably to band crossings along the
$\langle111\rangle$ directions.

Note that the system in the presence of an external applied magnetic
field, discussed in Ref.~\cite{moon2012}, is less symmetric. The
system's Hamiltonian at fixed $\mathbf{k}$, which we call
$\mathcal{H}_2(\mathbf{k})$, is only invariant under three-fold
rotations about $\mathbf{k}$ if both the magnetic field and
$\mathbf{k}$ point along the same $\langle111\rangle$ direction. For
$\mathbf{k}=\mathbf{0}$ the system has additionally inversion
symmetry, but all the representations of the symmetry group are
one-dimensional anyway, and there is a priori no degeneracy at
$\mathbf{k}=\mathbf{0}$. Away from $\mathbf{k}=\mathbf{0}$, any band
crossing is, again, accidental.

It is important to note that, although no crossings are required by
symmetry, once the crossings are found to happen, 
their properties are ``stable'' in the sense that {\em (i)} no
symmetry-preserving perturbation will remove them, {\em (ii)} the
dispersion along the crossings 
will remain linear, {\em (iii)} they
will not move away from the $\langle111\rangle$ axes.

\subsection{Couplings}

The long-range Coulomb interaction is described by introducing the
Hubbard-Stratonovich field, $\varphi$, which couples to the density of
fermions. 

The all-in all-out operator is represented by the time-reversal
symmetry breaking Ising field ($\phi$) corresponding to $J_x J_y J_z
+J_z J_y J_x$ in Luttinger's notation \cite{luttinger1956}.
In terms of the Gamma matrices, the order parameter is $\Gamma_{45} \sim J_x J_y J_z +J_z J_y J_x$. 
Thus, finally, the interaction part of the action is the ``vertex term'' given,
in real space and imaginary time, by 
\begin{eqnarray}
\mathcal{S}_{vertex} = \int d^3 x\, d {\tau}\, {\psi^\dagger} \left[i e\, \varphi + g\, \phi\, \Gamma_{45} \right]\psi,  
\end{eqnarray}
where $\psi$ is the four-component spinor field. Upon extending the
field space to $N$ flavors of fermions, this term becomes
\begin{eqnarray}
\mathcal{S}_{vertex} \rightarrow \frac{1}{\sqrt{N}}\int d^3 x\, d {\tau}\, {\psi^\dagger} \left[i e\, \varphi + g\, \phi\, \Gamma_{45} \right]\psi.
\end{eqnarray}

By appropriately transforming the Gamma matrices with transformations
not belonging to the cubic group, one may show that the signs of $c_{0,1,2}$ may always be taken
positive. Therefore,
throughout the paper we assume $c_{0,1,2}\geq0$. We also assume $c_0\leq
c_1/\sqrt{6}$, i.e.\ we assume the two sets of bands have opposite
curvatures in all directions at the Gamma point, or, in other words
that the Fermi energy goes through the band touching point.

\subsection{Green's function and self-energy conventions}

We use the following conventions for the boson Green's functions,
$\mathcal{G}_{b;\omega_n,\mathbf{k}}$ with $b=\phi,\varphi$, fermion
Green's function, $G_{\omega_n,\mathbf{k}}$, boson self-energies,
$\Sigma_b(\omega_n,\mathbf{k})$ and fermion self-energy, $\Sigma_f(\omega_n,\mathbf{k})$:
\begin{eqnarray}
\mathcal{G}_{\varphi;\omega_n,\mathbf{k}}&=&\langle\varphi_{-\mathbf{k}}\varphi_{\mathbf{k}}\rangle=\frac{1}{\mathbf{k}^2+\Sigma_\varphi(\mathbf{k})},\nonumber\\
\mathcal{G}_{\phi;\omega_n,\mathbf{k}}&=&\langle\phi_{-\omega_n,-\mathbf{k}}\phi_{\omega_n,\mathbf{k}}\rangle=\frac{1}{\mathbf{k}^2+\omega_n^2+r+\Sigma_\phi(\omega_n,\mathbf{k})},\nonumber\\
G^{\mu\nu}_{\omega_n,\mathbf{k}}&=&\langle\psi_{\omega_n,\mathbf{k}}^{\mu}{\psi_{\omega_n,\mathbf{k}}^{\nu}}^\dagger\rangle\nonumber\\
&=&\left[-i\alpha\omega_n+\mathcal{H}_0(\mathbf{k})+\Sigma_f(\omega_n,\mathbf{k})\right]^{-1},\nonumber
\end{eqnarray}
where $\mu,\nu=1,..,4$ (or $1,..,4N$) but are omitted throughout. The ``bare propagators'' are denoted with the
subscript or superscript ``$0$.''

\bigskip

\section{Asymptotic limits of the bosonic self-energies}
\label{sec:asymp}

We first evaluate the boson self-energies in the large-$N$ limit. They
are given by
\begin{eqnarray}
&&\Sigma_b(\omega_n,\mathbf{k})=\\
&&\;\;\frac{g_b^2}{N}\int_\Lambda\frac{d^3q}{(2\pi)^3}\int_{-\infty}^{+\infty}\frac{d\Omega_n}{2\pi}\,{\rm Tr}\left[G^0_{\Omega_n,\mathbf{q}}M_bG^0_{\Omega_n+\omega_n,\mathbf{q}+\mathbf{k}}M_b\right],\nonumber
\end{eqnarray}
where $g_\varphi=ie$, $g_\phi=g$, $M_\varphi=I$ and
$M_\phi=\Gamma_{45}$ ($I$ is the identity matrix).  Here the subscript $\Lambda$ in the $q$
integral indicates that an ultraviolet cutoff is required to keep
$\Sigma_b(0,\mathbf{0})$ finite.   This determines the (non-universal) location
of the QCP.  However, we seek the corrections to this term for
non-zero frequency and momenta, which are cutoff independent, and will 
be therefore obtained below without further discussion of $\Lambda$.
We will return later to the role of the cutoff when considering
fermionic self-energy terms, and treat it in more detail. The explicit
expression for $\Sigma_b(\omega_n,\mathbf{k})$ at $c_0\leq c_1/\sqrt{6}$ is
\begin{widetext}
\begin{equation}
\Sigma_b(\omega_n,\mathbf{k}) \label{eq:sigmabsmallc0}=\frac{-g_b^2}{\alpha}\sum_{\epsilon=\pm}\int_\Lambda\frac{d^3q}{(2\pi)^3}\left[\frac{E_{\mathbf{q},\mathbf{k}}^++E_{\mathbf{q},\mathbf{k}}^-+\epsilon\,2c_0\mathbf{q}\cdot\mathbf{k}}{\alpha^2\omega_n^2+\left(E_{\mathbf{q},\mathbf{k}}^++E_{\mathbf{q},\mathbf{k}}^-+\epsilon\,2c_0\mathbf{q}\cdot\mathbf{k}\right)^2}\right]
\left(1-\frac{F_{b;\mathbf{q},\mathbf{k}}}{E_{\mathbf{q},\mathbf{k}}^+E_{\mathbf{q},\mathbf{k}}^-}\right),\nonumber
\end{equation}
\end{widetext}
where $E_{\mathbf{q},\mathbf{k}}^\pm=E(\mathbf{q}\pm\mathbf{k}/2)$ and 
$F_{b;\mathbf{q},\mathbf{k}}=\sum_{a=1}^5(\varepsilon_a)^b\hat{c}_a^2d_a(\mathbf{q}-\mathbf{k}/2)d_a(\mathbf{q}+\mathbf{k}/2)$,
with $\varepsilon=(1\;1\;1\;-1\;-1)$ and $b=0$ (resp.\ $b=1$) for $b=\varphi$ (resp.\ $b=\phi$).
Note that $\Sigma_b$ is $O(1)$ (and not $O(1/N)$); mathematically this is because of the trace, which yields a factor of $N$.

As mentioned above, the boson self-energy $\Sigma_\phi(0,\mathbf{0})$ is finite but depends upon
the cutoff ($\Sigma_\phi(0,\mathbf{0})$ is proportional to $\Lambda$).
Again, this determines the
location of the QCP at $N=\infty$, and when we focus on the critical
theory, this zero-frequency zero-momentum contribution is exactly cancelled by the bare value of
$r$.  Hence we are left with the corrections at non-zero frequency and
momenta, which we isolate by considering the self-energy difference
$\Sigma_b(\omega_n,\mathbf{k})-\Sigma_b(0,\mathbf{0})$ (for
$b=\varphi$ the second term is zero by charge conservation).  This
difference is finite and cutoff independent.  In the
$c_{0,1}\rightarrow0$ limit, which will be the case in the critical
theory, the self-energy differences show logarithmic divergences,
i.e.\ contain $|\ln c_1/c_2\,|$. Conveniently, as mentioned in the
main text, the latter will act as a control parameter \cite{huh2008},
in addition to $N$, in the critical theory.

In the following, we thereby obtain the one-loop bosonic self energy, 
\begin{eqnarray}
&&\Sigma_b(\omega_n,\mathbf{k})-\Sigma_b(0,\mathbf{0})\label{eq:sigmabexpr}\\
&&\qquad=\frac{g_b^2}{\alpha}\left(|\mathbf{k}|f_b(\mathbf{\hat{k}})|\ln
  c_1/c_2\,|+\sqrt{|\omega_n|}\,C_b\right) \nonumber.
\end{eqnarray}
For future convenience, we take henceforth $c_2=1$ and denote $c=c_1$. It is
straightforward to obtain the coefficients of the frequency
dependences, $C_b$. Because $\Sigma_b$ is {\em larger} than the bare
term at $r=0$, which goes as $\mathbf{k}^2+\omega_n^2$, throughout
this work, we take $\mathcal{G}_b\rightarrow\Sigma_b^{-1}$, where
$\mathcal{G}_b$ is a full boson Green's function. Finally, note that we used an expansion in small $1/|\ln c_1/c_2\,|$
of $\Sigma_b^{-1}$, i.e.\ of the inverse of Eq.~(\ref{eq:sigmabexpr}), in some of the calculations.

By evaluating $\Sigma_b(\omega_n,\mathbf{0})-\Sigma_b(0,\mathbf{0})$, 
we find $C_\varphi=0$ and $C_\phi=1.33$ %hellonumber Cphi
taking $\alpha=1$, $c_1=0$ and $c_2=1$.
Note that in the $c_1/c_2 \rightarrow 0$ limit, the frequency
dependence is subdominant and the bosonic propagator becomes static.

We now extract the non-trivial logarithmic momentum dependence,
$f_b(\mathbf{\hat{k}})$.

\subsection{Coefficient of the logarithm}

As mentioned above, when $c_{0,1}=0$, to which the theory flows at the QCP, the
energy $E(\mathbf{k})$ and spectrum $\mathsf{E}_\pm(\mathbf{k})$ vanish for any
$\mathbf{k}\parallel\langle111\rangle$, which renders the self-energy
difference,  $\Sigma_b(\omega_n,\mathbf{k})-\Sigma_b(0,\mathbf{0})$,
{\em divergent}.  The appearance of a divergence is subtle:  for
general $\mathbf{k}$, the denominator in
Eq.~(\ref{eq:sigmabsmallc0}) appears relatively well-behaved since the singularity
occurs only when {\em both} $\mathbf{q}+ \mathbf{k}/2$ {\em and}
$\mathbf{q}- \mathbf{k}/2$ lie along a $\langle 111\rangle$ axis.  The
singularity actually arises from the regions of integration at large
$|\mathbf{q}|$ along these directions, where $|\mathbf{k}| \ll |\mathbf{q}|$, so that {\em both}
energies are small.  We analyze it below. In the limit $0\leq c_0 \ll c_1 \ll c_2=1$ (i.e. with
$c_1$ {\em nonzero} and small), which is the actual behavior in the 
RG {\em flows}, the divergence is removed, and the result is large in
$|\ln c_1/c_2\,|$.  In this subsection, we extract the leading result in this
limit.  Notably, in this limit, the result is independent of $c_0$, and
can be approximated by taking simply $c_0=0$.

To extract the coefficient of the logarithm, $f_b(\mathbf{\hat{k}})$,
we rotate to bases whose $x$-axes point along one of the
$\langle111\rangle$ directions, and make a change of variables such
that
\begin{equation}
\left\{\begin{array}{l}
\mathbf{\hat{e}}_1=(s_1,s_2,s_3)/\sqrt{3}\\
\mathbf{\hat{e}}_2=(0,s_2,-s_3)/\sqrt{2}\\
\mathbf{\hat{e}}_3=(-2s_1,s_2,s_3)/\sqrt{6}
\end{array}\right.
\;\mbox{and}\quad \mathbf{q}=\frac{Q}{c_1}\mathbf{\hat{e}}_1+u\mathbf{\hat{e}}_2+v\mathbf{\hat{e}}_3,
\end{equation}
where $s_i=\pm1$ (allows to span the eight $\langle111\rangle$
directions). This rewriting is chosen so that for $Q,u,v$ of $O(1)$,
the region near the $(s_1 s_2 s_3)$ ray is singled out.   The Jacobian of this coordinate transformation is
$\mathcal{J}_0=|s_1s_2s_3/c_1|$. Now,
we rewrite the functions involved in the integrand of the
self-energies, Eq.~(\ref{eq:sigmabsmallc0}), in these new coordinates,
and we obtain the leading asymptotic behavior of each such function at small
$c_1$.

For example, we find
\begin{equation}
{E_{\mathbf{q},\mathbf{k}}^\pm}\approx\frac{1}{c_1}\epsilon^\pm_{Q,u,v;k_1,k_2,k_3}
\quad\mbox{and}\quad
{F_{b;\mathbf{q},\mathbf{k}}}\approx\frac{1}{c_1^2}\gamma^b_{Q,u,v;k_1,k_2,k_3},
\end{equation}
where the $\epsilon_\pm$ and $\gamma^b$ ($b=\varphi,\phi$) are functions of
$\{Q,u,v,k_1,k_2,k_3\}$ (and of course of the $s_i$'s)
{\em only}.  We are then in a position to take the logarithmic
derivatives of the boson self-energies. A major simplification thereby
occurs: the frequency dependence drops out of
$\Sigma_b(\omega_n,\mathbf{k})-\Sigma_b(\omega_n,\mathbf{0})$. We find
\begin{eqnarray}
&&\frac{\alpha}{g_b^2}c_1\,\partial_{c_1}\left[\Sigma_b(\omega_n,\mathbf{k})-\Sigma_b(\omega_n,\mathbf{0})\right]\label{eq:fint}\\
&&\qquad=\sum_{s_1,s_2,s_3=\pm1}\int_{0}^{+\infty}\frac{dQ}{2\pi}\int_{-\infty}^{+\infty}\frac{du}{2\pi}
\int_{-\infty}^{+\infty}\frac{dv}{2\pi}\; \mathcal{K}_{s_1s_2s_3}^b\nonumber\\
&&\qquad=f_b(\mathbf{k})=|\mathbf{k}|f_b(\mathbf{\hat{k}}),
\end{eqnarray}
where 
\begin{eqnarray}
  \label{eq:mathcalK}
  \mathcal{K}^b_{s_1,s_2,s_3} & =& 9\sqrt{2}\,Q
  \Big[ \frac{3(a_0-h^b_0)}{a_0^{5/2}} 
   \\
   &&+\;
   \frac{2\left(h^b(a_++\sqrt{a_+}\sqrt{a_-}+a_-)-3a_+a_-\right)}{a_-^2a_+^{3/2}+a_+^2a_-^{3/2}}
   \Big]. \nonumber
\end{eqnarray}
In the above formula, we introduced several expressions:
\begin{eqnarray}
\kappa &=& s_1s_2k_xk_y + s_1s_3k_xk_z + s_2s_3k_yk_z\\
  h_{0}^\phi &=& 3\left[Q^2 - 2\left(u^2 +
      v^2\right)\right]\\
 a_0 = h_0^\varphi &=& 3\left[Q^2 + 2\left(u^2 + v^2\right)\right]\\
  h^\phi &=& h_0^\phi + \left(\mathbf{k}^2 -\kappa\right)\\
  h^\varphi &=& h_0^\varphi - \left(\mathbf{k}^2 -\kappa\right)\\
  a_\pm &=& a_0 + \left(\mathbf{k}^2 -\kappa \pm
    3\sqrt{2}u(s_2k_y-s_3k_z)\right. \nonumber\\
&&\left. \pm \sqrt{6}v(s_2k_y + s_3k_z - 2s_1k_x)\right),
\end{eqnarray}
where all the functions defined above, namely $h_0^b$, $a_0$, $h^b$,
$a_\pm$, and $\mathcal{K}^b$ ($b=\phi,\varphi$), are taken at
$\{Q,u,v,k_x,k_y,k_z\}$ (and are also functions of the
$s_i$'s although we have written the latter explicitly for
$\mathcal{K}^b$ only). Note that the integrations over $u$ and $v$ are taken
all the way from $-\infty$ to $+\infty$ although the sum over the
eight directions, $\sum_{s_1,s_2,s_3}$ is also taken.  This is
because, for non-zero
$c_1$, the $u,v$ integrations have a priori upper bounds of order
$Q/c_1$, which is taken to infinity.  In the present order of limits,
all contributions arise from regions of angular width of
order $c_1$ from the $\langle111\rangle$ rays.

The integrals, Eq.~(\ref{eq:fint}), are evaluated thanks to the Cuba
library, using the ``Cuhre'' routine \cite{hahn2005}.

\subsection{Approximation}

\begin{figure}[htbp]
\begin{center}
\includegraphics[width=2in]{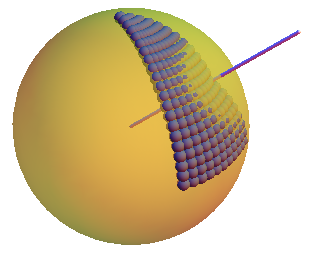}
\caption{Plot of $f_\phi(\mathbf{\hat{k}})/f_\phi\!\left(001\right)$. The line represents a $\langle111\rangle$ direction. The whole surface can be obtained from the plotted points by applying cubic symmetries (note that the set of plotted points is larger than the minimal set of points). The yellow surface is a sphere of radius $f_\phi\!\left(001\right)$.}
\label{fig:fplot}
\end{center}
\end{figure}

Since $f_b$ is very smooth (see Fig.~\ref{fig:fplot}), we approximate it by a low-order
polynomial of $\mathbf{k}$ in order to be able to take accurate
derivatives of $f_b$ as required to compute the flow of $c_1$ (and
$c_2$) -- see Sec.~\ref{sec:RGeqs}. Imposing cubic symmetry, the most
general polynomial to order six can take the form
\begin{equation}
\frac{1}{f_b(\mathbf{\hat{k}})}\approx m_1^b+m_2^b\left(\hat{k}_x^4+\hat{k}_y^4+\hat{k}_z^4\right)+m_3^b\,\hat{k}_x^2\hat{k}_y^2\hat{k}_z^2,
\end{equation}
%hellonumber m's
and fits with $m_1^\phi=2.356$, $m_2^\phi=-0.130$ and $m_3^\phi=4.136$ and $m_1^\varphi=-4.704$, $m_2^\varphi=0.264$ and $m_3^\varphi=-8.253$ provide excellent approximations: the square roots of the means of the squares are $R_\phi=0.0049$ and $R_\varphi=0.0049$, where $R_b=\frac{1}{N_{\rm pts}}\sqrt{\sum_{i=1}^{N_{\rm pts}}\frac{\left((1/f^b_i)-{\rm fit}^b_i\right)^2}{(1/f^b_i)^2}}$.

\section{RG equations}
\label{sec:RGeqs}

As discussed in the main text, twenty-four Feynman diagrams are necessary to
determine the RG equations: two boson self-energies, $\Sigma_b$, given
in Sec.~\ref{sec:asymp}, two fermion self-energies, $\Sigma_{f;b}$, and twenty
vertex corrections, the one-loop $\Xi_{b;(1);b'}$ (four) and the
two-loop $\Xi_{b;(2);b',b'',\eta}$
(sixteen, twelve of which either vanish identically or cancel out one
another), with $b,b',b''=\varphi,\phi$ and $\eta=\pm1$. The notation is expected to be
transparent, %with $\Sigma_f=\sum_{b=\phi,\varphi}\Sigma_{f;b}$,
%$\Xi_b=\Xi_{b;(1)}+\Xi_{b;(2)}$,
%$\Xi_{b;(1)}=\sum_{b'=\phi,\varphi}\Xi_{b;(1);b'}$ and
%$\Xi_{b;(2);b',b'',\eta}=\sum_{b',b''=\phi,\varphi;\eta=\pm}\Xi_{b;(2);b',b'',\eta}$,
and the expressions can be read off in Eqs.~(\ref{eq:sigmafeq}--\ref{eq:xi2}). We proceed like in
Refs.~\cite{vojta2000a,huh2008}, i.e.\ we find the
corrections to the parameters of the theory by evaluating the former
when a small change in the cutoff is applied. It physically
corresponds to integrating out modes to keep the rescaled cutoff unchanged. In
practice, we {\em (i)} use
soft momentum-cutoffs for the integrals, implemented by the use of a
rapidly decaying function
$|\mathbf{q}|/\Lambda\mapsto\mathcal{F}(|\mathbf{q}|/\Lambda)$, with
e.g.\ $\mathcal{F}$ belonging to the function space $\mathcal{L}^2(\mathbf{R})$, {\em (ii)} compute the
logarithmic derivatives with respect to the cutoff $\Lambda$ of the
fermion self-energy and vertices, {\em (iii)} identify the
appropriate coefficients of the Taylor expansion (in $k_i$ and
$\omega_n$) of the result. The choice of a soft cutoff is fairly
arbitrary, but helps to avoid spurious singularities induced by
``ringing'' at the spectral edge.  The derivative with respect to
$\Lambda$ serves to extract the {\em incremental} change in the band
parameters due to a small change of cutoff, as in the Wilsonian view
of RG.  The momentum and frequency expansion allows identification of
the renormalization of each term of the Hamiltonian independently.

\subsection{Diagram expressions}

The fermion self-energy is 
\begin{eqnarray}
&&\Sigma_{f}(\omega_n,\mathbf{k})
=\sum_{b=\varphi,\phi}\frac{-g_b^2}{N} \label{eq:sigmafeq}\\
&&\quad\times\int\frac{d^3q}{(2\pi)^3}\int_{-\infty}^{+\infty}\frac{d\Omega_n}{(2\pi)}\frac{M_bG^0_{\Omega_n,\mathbf{q}}M_b
\mathcal{F}\left(\frac{|\mathbf{q}|}{\Lambda}\right) \mathcal{F}\left(\frac{|\mathbf{k}-\mathbf{q}|}{\Lambda}\right)}{\Sigma_b({\omega_n-\Omega_n,\mathbf{k}-\mathbf{q}})-\Sigma_b(0,\mathbf{0})},\nonumber
\end{eqnarray}
where two cutoff functions $\mathcal{F}$ are present because both
fermion lines in the self-energy should be cutoff, i.e. the momenta of all
the electrons in the theory are taken within the cutoff.  Similarly, the vertex
corrections {\em at zero external momenta and frequencies} are
$\Xi_b^0=\Xi_{b;(1)}^0+\Xi_{b;(2)}^0$, with
\begin{eqnarray}
\label{eq:xi1}
&&\Xi_{b;(1)}^0=\sum_{b'=\varphi,\phi}\frac{g_b g_{b'}^2}{N^{3/2}}\\
&&\quad\times\int\frac{d^3q}{(2\pi)^3}\int_{-\infty}^{+\infty}\frac{d\Omega_n}{2\pi}\frac{M_{b'}G^0_{\Omega_n,\mathbf{q}}M_bG^0_{\Omega_n,\mathbf{q}}M_{b'}\mathcal{F}^2\left(\frac{|\mathbf{q}|}{\Lambda}\right)}{\Sigma_{b'}({\Omega_n,\mathbf{q}})-\Sigma_{b'}({0,\mathbf{0}})},\nonumber
\end{eqnarray}
and
\begin{widetext}
\begin{eqnarray}
\label{eq:xi2}
&&\Xi_{b;(2)}^0=-\sum_{b',b''=\varphi,\phi;\eta=\pm}\frac{g_b  g_{b'}^2g_{b''}^2}{N^{5/2}}\int\frac{d^3q_1}{(2\pi)^3}\int\frac{d^3q_2}{(2\pi)^3}\int_{-\infty}^{+\infty}\frac{d\Omega_{n,1}}{2\pi}\int_{-\infty}^{+\infty}\frac{d\Omega_{n,2}}{2\pi}\\
&&\qquad\qquad\times\frac{M_{b'}G^0_{\Omega_{n,2},\mathbf{q}_2}M_{b''}{\rm    Tr}\left\{G^0_{\Omega_{n,1},\mathbf{q}_1}M_{b'}G^0_{\Omega_{n,1}+\eta\Omega_{n,2},\mathbf{q}_1+\eta\mathbf{q}_2}M_{b''}G^0_{\Omega_{n,1},\mathbf{q}_1}M_b\right\}}{\left[\Sigma_{b'}({\Omega_{n,2},\mathbf{q}_2})-\Sigma_{b'}({0,\mathbf{0}})\right]\left[\Sigma_{b''}({\Omega_{n,2},\mathbf{q}_2})-\Sigma_{b''}({0,\mathbf{0}})\right]}\mathcal{F}\left(\frac{|\mathbf{q}_1|}{\Lambda}\right) \mathcal{F}\left(\frac{|\mathbf{q}_2|}{\Lambda}\right) \mathcal{F}\left(\frac{|\mathbf{q}_1+\eta\mathbf{q}_2|}{\Lambda}\right).\nonumber
\end{eqnarray}
\end{widetext}
All other diagrams are smaller in a $1/N$ expansion. By using for
example
$\partial_{-i\alpha\omega_n}G^0_{\omega_n,\mathbf{k}}=-(G^0_{\omega_n,\mathbf{k}})^2$,
one can show that the two-loop diagrams, $\Xi_{b;(2);b',b'',\eta}^0$, with
identical internal boson propagators ($b'=b''$) cancel out one another
upon performing the sum over $\eta=\pm1$ (and even vanish identically
in the case $b=\phi$). The remaining two-loop diagrams correcting the
Coulomb vertex ($b=\varphi$ and $b'\neq b''$) can also be shown to
vanish, for example by noticing that only the $b'=b''$ diagrams can
renormalize $g$. Therefore, only four two loop diagrams (those with $b=\phi$
and $b'\neq b''$), shown in Fig.~3 of the main text, need be
calculated. Careful observation shows all contributions are equal, and
an explicit calculation yields a finite integral, which converges to a
nonzero value multiplied by $g/(N^{3/2}|\ln c_1/c_2\,|^2)$.  This is
actually subdominant (for $c_1/c_2 \ll 1$) to the contribution from
the one loop vertex correction, although it is of the same order in $1/N$.

\subsection{Flow equations}

We find the following RG flow equations (``beta-functions''). The flow
of $\alpha$, the coefficient of the frequency in the fermion self-energy, is
\begin{equation}
\frac{\partial_\ell
\alpha}{\alpha}=3-2\Delta_\psi+\frac{1}{\alpha}\left.\left(\partial_{-i\omega_n}\left[D_\Lambda\Sigma_f^0\right]\right)\right|_{\omega_n=0,\mathbf{k}=\mathbf{0}}.
\label{eq:betaalpha}
\end{equation}
where $D_\Lambda=\Lambda\frac{d}{d\Lambda}$. As usual, the last term
of the right-hand-side corresponds in the RG procedure to the
``rescaling'' (or integration of momenta), while the other terms
correspond to the ``renormalization'' \cite{kardar2007}. The ``anisotropic''
coupling of the fermions to the bosons leads to ``anisotropic''
corrections to the coefficients of the fermion Hamiltonian:
\begin{equation}
\frac{\partial_\ell
  c_j}{c_j}=z+1-2\Delta_\psi+\frac{\left(\delta\Sigma_f\right)_j^0}{c_j},\qquad
j=0,1,2,
\label{eq:betac0}
\end{equation}
where
\begin{equation}
\left(\delta\Sigma_f\right)_j^0=
\begin{cases}
\frac{1}{2}\left.\left(\partial_{k_x,k_x}^2\left[D_\Lambda\Sigma_f^0\right]\right)\right|_{\omega_n=0,\mathbf{k}=\mathbf{0}}
& \mbox{for }j=0\\
\sqrt{2}\left.\left(\partial_{k_x,k_y}^2\left[D_\Lambda\Sigma_f^1\right]\right)\right|_{\omega_n=0,\mathbf{k}=\mathbf{0}}&
\mbox{for }j=1\\
\sqrt{2}\left.\left(\partial_{k_x,k_x}^2\left[D_\Lambda\Sigma_f^4\right]\right)\right|_{\omega_n=0,\mathbf{k}=\mathbf{0}}& \mbox{for }j=2
\end{cases}.
\label{eq:dLambdaetc}
\end{equation}
The RG equations for the coupling constants are simply:
\begin{equation}
\frac{\partial_\ell
  g_b}{g_b}=z+3-\Delta_\phi-2\Delta_\psi+M_b^{-1}\frac{\left[D_\Lambda\Xi_b^0\right]}{g_b/\sqrt{N}},
\label{eq:betag}
\end{equation}
for $g_\phi=g,g_\varphi=ie$ and $M_\phi=\Gamma_{45}$, $M_\varphi=I$. The right-hand-sides of the equations eventually involve angular
integrals that can be performed numerically, and which are obtained using the
identities:
\begin{equation}
\begin{cases}
\int_0^\infty dq\;\frac{1}{q}\Lambda\frac{d}{d\Lambda}\left[\mathcal{F}^2(q/\Lambda)\right]=1\\
\int_0^\infty dq\; \Lambda\frac{d}{d\Lambda}\left[\frac{\mathcal{F}(q/\Lambda)\mathcal{F}'(q/\Lambda)}{\Lambda}\right]=0\\
\int_0^\infty dq\;
q\Lambda\frac{d}{d\Lambda}\left[\frac{\mathcal{F}(q/\Lambda)\mathcal{F}''(q/\Lambda)}{\Lambda^2}\right]
=0\\
\int_0^\infty dq_1 \frac{\Lambda}{q_1}\frac{d}{d\Lambda}\left[\mathcal{F}(q_1/\Lambda) \mathcal{F}(q_1\tilde{q}_2/\Lambda) \mathcal{F}(q_1(1+\tilde{q}_2)/\Lambda)\right]=1
\end{cases}
\end{equation}
(for any $\tilde{q}_2$), since $\mathcal{F}(0)=1$ and $\mathcal{F}(+\infty)=0$.

In practice, to calculate the flows of $\alpha$ and $c_0$, from
Eqs.~(\ref{eq:betaalpha}) and (\ref{eq:betac0}) with $j=0$, we
shift the 
internal momentum in the integrands of $\Sigma_f$ (see
Eq.~(\ref{eq:sigmafeq})), i.e.\ 
$\mathbf{q}\rightarrow\mathbf{q}+\mathbf{k}$. As a result, the derivatives
with respect to the frequency $\omega_n$ or momenta $k_i$ involve the
fermionic part of the integrands. Proceeding otherwise to obtain the
equation for $c_0$ leads to a divergent integral. For the flow of $c_2$, where the derivatives with respect to either
part of the integral converge, we have checked that both ``methods''
give the same result. The integrals from the vertex corrections
converge, in particular, we find the double integrals in $[D_\Lambda \Xi_{b;(2)}^0]$
are subdominant (equal to a finite number times $1/|\ln c_1/c_2\,|^2$,
the latter factor
coming solely from the two inverse
boson propagators), even upon taking $c_{0,1}=0$ directly in $G^0$.

\subsection{Details of the flows of $c_1$ and $c_2$}

Because the results are crucial to the physics, we give the details of the
calculation of the beta functions for $c_1$ and $c_2$. Applying the
derivatives in Eq.~(\ref{eq:dLambdaetc}) with $j=1,2$ to the ``boson parts'' of the integrand
in the self-energies using the approximations discussed in
Sec.~\ref{sec:asymp}, and expanding $\Sigma_b^{-1}$ in small $1/|\ln c_1/c_2\,|$, we find:
\begin{widetext}
\begin{eqnarray}
\frac{\left(\delta\Sigma_f\right)_1^0}{c_1}&=&-\frac{\sqrt{2}}{8\pi|\ln c_1/c_2\,|N}\int\frac{d\mathbf{\hat{q}}}{(2\pi)^2}\frac{d_1(\mathbf{\hat{q}})}{E_\mathbf{\hat{q}}}\left\{(m_1^\varphi+m_1^\phi)\mathcal{N}_{1,1}+(m_2^\varphi+m_2^\phi)\mathcal{N}_{1,2}+(m_3^\varphi+m_3^\phi)\mathcal{N}_{1,3}\right\}\\
\frac{\left(\delta\Sigma_f\right)_2^0}{c_2}&=&-\frac{\sqrt{2} }{8\pi|\ln c_1/c_2\,|N}\int\frac{d\mathbf{\hat{q}}}{(2\pi)^2}\frac{d_4(\mathbf{\hat{q}})}{E_\mathbf{\hat{q}}}\left\{(m_1^\varphi-m_1^\phi)\mathcal{N}_{2,1}+(m_2^\varphi-m_2^\phi)\mathcal{N}_{2,2}+(m_3^\varphi-m_3^\phi)\mathcal{N}_{2,3}\right\},
\end{eqnarray}
\end{widetext}
where 
\begin{eqnarray}
\mathcal{N}_{1,1}&=&3\hat{q}_x\hat{q}_y\\
\mathcal{N}_{1,2}&=&5 \hat{q}_x \hat{q}_y \left( -8
  \hat{q}_x^2\hat{q}_y^2-4 \hat{q}_x^2\hat{q}_z^2-4 \hat{q}_y^2
  \hat{q}_z^2\right.\nonumber\\
&&\left.+3 \hat{q}_x^4+3 \hat{q}_y^4+7 \hat{q}_z^4\right)\\
\mathcal{N}_{1,3}&=&-\hat{q}_x \hat{q}_y \hat{q}_z^2 \left(6
  \hat{q}_x^2\hat{q}_z^2-43 \hat{q}_x^2\hat{q}_y^2+6 \hat{q}_y^2
  \hat{q}_z^2\right.\nonumber\\
&&\left.+10 \hat{q}_x^4+10 \hat{q}_y^4-4 \hat{q}_z^4\right)\\
\mathcal{N}_{2,1}&=&2\hat{q}_x^2-\hat{q}_y^2-\hat{q}_z^2\\
\mathcal{N}_{2,2}&=&24\hat{q}_x^2\hat{q}_y^2\hat{q}_z^2 -
21\hat{q}_x^4 \left(\hat{q}_y^2+\hat{q}_z^2\right)+\hat{q}_y^4
\left(42 \hat{q}_x^2 -5\hat{q}_z^2\right)\nonumber\\
&&+\hat{q}_z^4 \left(42 \hat{q}_x^2 -5\hat{q}_y^2\right)+2 \hat{q}_x^6-5\hat{q}_y^6-5\hat{q}_z^6\\
\mathcal{N}_{2,3}&=&\hat{q}_y^2 \hat{q}_z^2 \left(-31 \hat{q}_x^2
  \hat{q}_y^2-31 \hat{q}_x^2
  \hat{q}_z^2+4\hat{q}_y^2\hat{q}_z^2\right.\nonumber\\
&&\left. +30 \hat{q}_x^4+2\hat{q}_y^4+2\hat{q}_z^4\right).
\end{eqnarray}
The relative signs of the terms coming from $\Sigma_\phi$ originate
from the ``opposite'' commutation relations of $\Gamma_{1,2,3}$ and
$\Gamma_{4,5}$ with $\Gamma_{45}$, i.e.\
$[\Gamma_a,\Gamma_{45}]=0$ for $a=1,2,3$ and
$\{\Gamma_a,\Gamma_{45}\}=0$ for $a=4,5$. Note that this is true
before implementing any approximation or assumption on the magnitude
of $c_1/c_2$. If $e=0$, it is then obvious that
the flows of $c_1$ and $c_2$ will take different directions, i.e. that
the ratio $c_1/c_2$ will be either relevant or irrelevant, or in other
words, will flow either to infinity or zero. Hence a calculation
taking $c_1/c_2$ large or small from the beginning is for sure
valid. We find that $c_1/c_2\rightarrow0$ occurs for $e=0$ (see
below). When $e\neq0$, the situation is not as clear-cut, but taking
$c_1/c_2$ small, as when $e=0$, proves to be self-consistent as shown
below. We can also justify it {\em a posteriori} as
follows. $c_1/c_2\rightarrow+\infty$ would lead to a situation where
the coupling term $\phi\psi^\dagger\Gamma_{45}\psi$ commutes with the
bare Hamiltonian at the critical point, hence {\em removing} all
fluctuations due to the coupling to the order
parameter, which is supposed to drive the transition through the
fluctuations it induces. Such a choice seems therefore unreasonable. The
situation where $c_1/c_2\rightarrow c^*$, a fixed constant, although
perhaps seemingly more reasonable, would imply the existence of a
universal ratio, when none seems to be natural. Hence, the limit
$c_1/c_2\rightarrow0$ seems to be the only reasonable limit to be
taken. $c_0/c_1\rightarrow0$ is also consistent.

\subsection{Exponents}

Keeping $\alpha, c_2, g$ and $e$ constant, i.e.\ setting the corresponding flow equations to zero, the dynamical critical exponent and the field dimensions are
\begin{eqnarray}
&&z= 2 -\frac{a_z}{N|\ln c_1/c_2\,|}, \qquad \Delta_{\psi} = \frac{3}{2}+\frac{a_\psi}{N|\ln c_1/c_2\,|^2},\nonumber\\
&&\Delta_{\phi} = \frac{3}{2} + \left[\frac{1}{2} +\frac{a_{\phi}}{N|\ln
    c_1/c_2\,|}\right],\\
&&\Delta_{\varphi} = \frac{3}{2} + \left[\frac{1}{2} -\frac{a_{\varphi}}{N|\ln c_1/c_2\,|}\right],  \nonumber
\end{eqnarray}
%hellonumbers az, apsi, aphi, avarphi
where $a_{z}=0.063$, $a_{\psi} =0.143$,
$a_{\phi}=0.255$, and $a_\varphi=0.063$. The
anomalous dimensions are then simply $\delta z=a_z/(N|\ln
c_1/c_2\,|)$, $\eta_\psi=2a_\psi/(N|\ln c_1/c_2\,|^2)$,
$\eta_\phi=1+2a_\phi/(N|\ln c_1/c_2\,|)$ and
$\eta_\varphi=1-2a_\varphi/(N|\ln c_1/c_2\,|)$, as given in the main text.

\subsection{Solutions to the flow equations}

Finally, we obtain
\begin{eqnarray}
&&\partial_\ell \left(\frac{c_1}{c_2}\right)
=-\frac{c_1}{c_2}\frac{Y}{N|\ln c_1/c_2\,|},\\
&&\partial_\ell \left(\frac{c_0}{c_1}\right) =-\frac{c_0}{c_1}\frac{W}{N|\ln c_1/c_2\,|},
\label{eq:c1c2c0c1flows}
\end{eqnarray}
%hellonumbers Y, W
with $Y=0.020$ and $W=0.043$.  These equations are solved analytically by
\begin{equation}
\left(c_1/c_2\right)(\ell)=e^{-\frac{\upsilon_0}{\sqrt{N}}\sqrt{\ell+\ell_0}},\quad
\left(c_0/c_1\right)(\ell)=\Upsilon_0e^{-\frac{\upsilon_0'}{\sqrt{N}}\sqrt{\ell+\ell_0}},
\end{equation}
where $\upsilon_0=\sqrt{2Y}$ and $\upsilon_0'=\sqrt{2}W/\sqrt{Y}$, and
where $\ell_0$ and $\Upsilon_0$ are constants which depend on $c_{0,1,2}(\ell=0)$.

Note that, as mentioned in the main text, {\em in the absence of Coulomb
interactions}, we find %hellonumbers upsilon0nocoulomb, upsilonprimenocoulomb
$(c_1/c_2)(\ell)=e^{-\frac{0.359}{\sqrt{N}}\sqrt{\ell+\ell_1}}$ and
$(c_0/c_1)(\ell)\propto e^{\frac{0.240}{\sqrt{N}}\sqrt{\ell+\ell_1}}$ ($\ell_1$ is
a constant), i.e.\ $c_0/c_1$
is found to be a {\em relevant} parameter in that case. The latter
means that, eventually, $c_0$ reaches $c_1/\sqrt{6}$, point at which
Fermi surfaces start to develop, rendering our theory invalid and the
heretofore studied critical point unstable. This
would correspond to a Lifshitz transition.

\section{Physical quantities}

We are now in a position to calculate the behavior of some physical
quantities. We first extract the critical exponent of the correlation
length. The associated RG flow is
\begin{equation}
\frac{\partial_\ell r}{r} = z+3-2\Delta_\phi,\quad\mbox{i.e.}\quad
\partial_\ell r=\nu^{-1}(\ell)r,
\end{equation}
with
\begin{equation}
\nu^{-1}(\ell)=1-\frac{2a_\phi+a_z}{N|\ln c_1/c_2\,|}.
\end{equation}
So
\begin{equation}
\partial_\ell r=\left(1-\frac{A}{\sqrt{N}\sqrt{\ell+\ell_0}}\right)r
,\quad\mbox{with}\quad A=\frac{2a_\phi+a_z}{\upsilon_0},
\end{equation}
%hellonumbers A
i.e. $A=2.836$, which is solved into
\begin{equation}
r(\ell)=r_0 e^{\ell-\frac{2A}{\sqrt{N}}\sqrt{\ell+\ell_0}},
\label{eq:rflow}
\end{equation}
where $r_0$ is a constant which depends on $r(\ell=0)$. We can easily invert $r=r(\ell)$ to $\ell=\ell(r)$ by taking the
log of Eq.~(\ref{eq:rflow}), squaring both sides and solving the quadratic equation. We get:
\begin{eqnarray}
\ell
&\approx&\ln\frac{r}{r_0}+\frac{2A}{\sqrt{N}}\sqrt{\ln\frac{r}{r_0}+\ell_0},
\label{eq:lfnofr}
\end{eqnarray}
where we have kept only terms to order $1/\sqrt{N}$.

\subsection{Order parameter exponent}

We first extract the exponent $\beta$ and its logarithmic correction, i.e. how $\langle\phi\rangle$ behaves with
$r$. We write 
\begin{equation}
\frac{\phi(\ell+d\ell)-\phi(\ell)}{\phi(\ell+d\ell)}\approx d\ell\Delta_\phi(\ell),
\end{equation}
and integrate both sides from $0$ to $\ell$. Using Eq.~(\ref{eq:lfnofr}), we obtain
\begin{eqnarray}
\frac{\phi}{\phi_0}&\sim&\left(\frac{r}{r_0}\right)^2\exp\left[2\,\frac{5a_\phi+2a_z}{\upsilon_0\sqrt{N}}\sqrt{\ln\frac{r}{r_0}+\ell_0}\right]\\
&&\qquad\qquad\qquad\qquad\qquad\qquad\times\exp\left[\frac{-2a_\phi\sqrt{\ell_0}}{\upsilon_0\sqrt{N}}\right],\nonumber
\end{eqnarray}
%hellonumbers 2(5aphi+2az)/upsilon0, 2aphi/upsilon0
with $2\,\frac{5a_\phi+2a_z}{\upsilon_0}=13.867$ and
$2a_\phi/\upsilon_0=2.523$ (in the main text, we absorbed $\ell_0$ in
the definition of $r_0$). Contrary to more conventional problems,
like the usual Ising model,
where $\beta_{\rm Ising}=1/2$ in three spatial dimensions, the bosonic order parameter here grows very slowly
as one moves away from the critical point on the ordered side of the
transition. This can be seen to be due to the massive fluctuations of
the boson field due to the strong coupling to the fermions.

\subsection{Specific heat}

At the critical point (or in the quantum critical region), temperature
is the only relevant parameter, so thermal properties receive
intriguing corrections in our critical theory. Since the fermion is
well-defined ($\eta_{\psi} \rightarrow 0$), the thermal average of the energy  is 
\begin{equation}
\langle \mathsf{E}\rangle = \sum_{i=\pm,\mathbf{k}}\langle
n_{i,\mathbf{k}}\rangle\mathsf{E}_i(\mathbf{k}) = \sum_{i,\mathbf{k}}\frac{2}{e^{\beta\mathsf{E}_{i}(\mathbf{k})}+1}\mathsf{E}_{i}(\mathbf{k}),
\end{equation}
with, for $\langle\phi\rangle=0$, $c_0=0$ and $c_2=1$,
$\mathsf{E}_{\pm}(\mathbf{k})=\pm\frac{\mathbf{k}^2}{\sqrt{6}}\sqrt{1+3(c_1^2-1)w^2_{\mathbf{\hat{k}}}}$
where
$w_{\mathbf{\hat{k}}}=\hat{k}_x^2\hat{k}_y^2+\hat{k}_x^2\hat{k}_z^2+\hat{k}_y^2\hat{k}_z^2$. 
To lowest order, we find
\begin{eqnarray}
C_V=\partial_T\langle \mathsf{E}\rangle&\approx&\frac{15 (4-\sqrt{2}) 6^{3/4}\sqrt{\pi}}{16}\zeta(5/2)
\frac{T^{3/2}}{c_1^{3/2}}\\
&\approx&22.1\exp\left[\frac{3\upsilon_0}{2\sqrt{N}\sqrt{z}}\sqrt{\ln\frac{T_0}{T}}\right]T^{3/2},\nonumber
\end{eqnarray}
where %hellonumbers 3upsilon0/(2sqrt(z))
$3\upsilon_0/(2\sqrt{z})\approx0.215$ (we use $z\approx2$). To obtain the last line, we used the approximation
$e^\ell=\left(T(\ell)/T_0\right)^{1/z}$ and thereby solved the RG equation of $c_1$ in terms of temperature. 
The logarithmic correction to the $T^{3/2}$
law is a signature of the fact that $c_1$ becomes scale
(temperature)-dependent in the quantum critical region.

\section{Mean-field theory}

In this section we consider the behavior in the ordered phase
according to na\"ive mean field theory, i.e. a saddle-point evaluation
of the $\varphi$ and $\phi$ integrals.  The former saddle point is
simply $\varphi=0$, i.e.\ there are no effects of the long-range
Coulomb interactions at the mean field level.  The saddle point value of $\phi$
is non-zero in the antiferromagnetic phase.  It is governed by the effective action
which consists of the bare one (second line of Eq.~(1) of the main
text) {\em plus} the contribution obtained by integrating out the
fermions. 

The fermionic contribution to the effective action, for constant
$\phi$, is simply the space-time integral of the total ground state
energy density of the electrons.  This is obtained by summing up the
energy of occupied single-particle states.  

In the saddle point approximation, the Hamiltonian density of the
fermions is
\begin{equation}
\mathcal{H}^\psi_{\rm MF}[\phi]=c_0\mathbf{k}^2+\sum_{a=1}^5\hat{c}_ad_a(\mathbf{k})\Gamma_a+\phi\Gamma_{45},
\end{equation}
and we therefore have the ground state energy density
\begin{equation}
\mathcal{E}_{\rm
  MF}^\psi\left[\phi\right]=\sum_{\alpha=1}^2\int_\Lambda
\frac{d^3k}{(2\pi)^3}\, \mathsf{E}_\mathbf{k}^\alpha[\phi],
\label{eq:gsen}
\end{equation}
($\mathsf{E}_\mathbf{k}^{1,2}$ are the single-particle lowest-energy
bands, with
$\mathsf{E}^{1,2}_\mathbf{k}[\phi=0]=\mathsf{E}_-(\mathbf{k})$).  Here,
by diagonalizing $\mathcal{H}_{\rm MF}^\psi[\phi]$, we obtain
\begin{eqnarray}
&&\mathsf{E}^{1,2,3,4}_{\mathbf{k}}[\phi]=c_0\mathbf{k}^2\\
&&\quad\pm\frac{1}{\sqrt{6}}\sqrt{c_2^2\mathbf{k}^4+2(c_1^2-c_2^2)w_{\mathbf{\hat{k}}}^2\mathbf{k}^4\pm6\sqrt{2}c_1\mathbf{k}^2w_{\mathbf{\hat{k}}}\phi+6\phi^2}\nonumber,
\end{eqnarray}
where we define $\mathbf{k}^4=(\mathbf{k}^2)^2$, and where $1,2,3,4$
correspond to the signs $\{--,-+,+-,++\}$, respectively. 

From scaling, $\mathsf{E}_\mathbf{k}^{1,2} \sim \mathbf{k}^2$, and hence, from
Eq.~(\ref{eq:gsen}), one expects that the singular scaling contributions
to the effective action behave as $\mathcal{E}_{\rm
  MF}^\psi \sim |\mathbf{k}|^5 \sim |\phi|^{5/2}$, where we used $\phi \sim \mathbf{k}^2$,
which follows dimensionally from $\mathcal{H}_{\rm MF}^\psi$.  This
describes only the singular contributions.  Since $\mathcal{E}_{\rm
  MF}^\psi$ is an even function of $\phi$, we expect it to contain
constant and quadratic terms as well (which are cutoff-dependent).
Indeed one can verify by direct expansion in $\phi$ that the 
integrals which arise from Eq.~(\ref{eq:gsen}) as coefficients of unity
and $\phi^2$ are finite, but if one proceeds to the following order, the
coefficient of $\phi^4$ is divergent.  This is due to the presence of
the $t |\phi|^{5/2}$ term.  

To extract the coefficient $t$, we take three derivatives of $\mathcal{E}_{\rm MF}^\psi[\phi]$ with respect to
$\phi$. We find an integral 
whose integrand goes as $1/|\mathbf{k}|^6$ at large $|\mathbf{k}|$, so
that the result is integrable in that region.  One then simply rescales
$\mathbf{k} \rightarrow \mathbf{k}/\sqrt{|\phi|}$, and takes the limit
of small $\phi$ (i.e.\ $\Lambda/\sqrt{\phi}\rightarrow+\infty$).  This makes
the singular behavior explicit, and in this limit we find 
%hellonumbers
$\partial_{\phi,\phi,\phi}^3 \mathcal{E}_{\rm
  MF}^{\psi}\left[\phi\right]=1.079/\sqrt{\phi}$,
i.e. $t=1.079\times\frac{8}{15}=0.575$, where the coefficient was
determined by a numerical integration taking the fixed-point values $c_0=c_1=0$. Therefore, $t|\phi|^{5/2}$
is indeed the lowest-order nonanalytical term. Hence, putting
everything together, and looking at the boson action with the
fermions integrated out, we have:
\begin{eqnarray}
\mathcal{S}_{\rm MF}\left[\phi\right]&=&V\int
d\tau\left[r\phi^2+\mathcal{E}^\psi_{\rm MF}[\phi]\right]\\
&\sim&V\int
d\tau\left[r'\phi^2+t|\phi|^{5/2}\right],
\end{eqnarray}
all other terms being irrelevant. Above, $r'$ includes the $\phi^2$
terms in $\mathcal{E}^\psi_{\rm MF}[\phi]$.   Most importantly, we
obtained positive $t>0$, so that when $r'<0$, a stable minimum action
configuration exists, describing a continuous --but unconventional--
transition at the mean field level.

\end{document}